\documentclass[twocolumn,preprintnumbers,nofootinbib,prl]{revtex4}
\usepackage{amsmath,amssymb,bm,graphicx,epsfig}
\begin{document}

\preprint{MZ-TH/10-45, EFI Preprint 10-30} 

\title{\boldmath Long-Distance Dominance of the CP Asymmetry in $\bar B\to X_{s,d}\,\gamma$ Decays}

\author{Michael Benzke$^a$, Seung J.~Lee$^b$, Matthias Neubert$^a$, and Gil Paz$^c$} 

\affiliation{${}^a$Institut f\"ur Physik (THEP), Johannes Gutenberg-Universit\"at, 
D-55099 Mainz, Germany\\
${}^b$Department of Particle Physics, Weizmann Institute of Science, 
Rehovot 76100, Israel\\
${}^c$Enrico Fermi Institute, University of Chicago, 
Chicago, IL 60637, U.S.A.}

\date{December 16, 2010}

\begin{abstract}
\noindent
We show that in the Standard Model the parametrically leading (by a factor $1/\alpha_s$) contribution to the inclusive CP asymmetry in $\bar B\to X_{s,d}\,\gamma$ decays arises from a long-distance effect in the interference of the electro-magnetic dipole amplitude with the amplitude for an up-quark penguin transition accompanied by soft gluon emission. Using model estimates for the associated hadronic parameter $\tilde\Lambda_{17}^u$, we predict a value in the range $-0.6\%<{\cal A}_{X_s\gamma}^{\rm SM}<2.8\%$. In view of current experimental data, a future precision measurement of the flavor-averaged CP asymmetry would signal the presence of new physics only if a value below $-2\%$ was found. A cleaner probe of new physics is offered by the difference of the CP asymmetries in charged versus neutral $\bar B\to X_s\gamma$ decays.
\end{abstract}

\maketitle

The radiative decays $\bar B\to X_s\gamma$ and $\bar B\to X_d\gamma$ are important for testing the Standard Model (SM) and probing its possible extensions. Both the CP-averaged branching ratios and the CP asymmetries are useful in this context. Thanks to a vigorous effort, the experimental error on the $\bar B\to X_s\gamma$ branching ratio has been reduced below 10\%, which is close to the (irreducible) theoretical uncertainty in the calculation of this observable. For the CP asymmetry, on the other hand, the experimental error is thought to be much larger than the theoretical one. The current world average is \cite{TheHeavyFlavorAveragingGroup:2010qj}
\begin{equation}\label{data}
   {\cal A}_{X_s\gamma} 
   = \frac{\Gamma(\bar B\to X_s\gamma)-\Gamma(B\to X_{\bar s}\gamma)}%
          {\Gamma(\bar B\to X_s\gamma)+\Gamma(B\to X_{\bar s}\gamma)} 
   = - (1.2\pm 2.8)\% \,.
\end{equation}
It is widely believed that the theoretical prediction for the CP asymmetry in the SM is short-distance dominated, leading to a tiny value of about 0.5\% as a result of a combination of perturbative, CKM, and GIM suppression \cite{Soares:1991te,Kagan:1998bh,Ali:1998rr}. A dedicated analysis finds ${\cal A}_{X_s\gamma}^{\rm SM}=(0.44_{\,-\,0.10}^{\,+\,0.15}\pm 0.03_{\,-\,0.09}^{\,+\,0.19})\%$ \cite{Hurth:2003dk}, where the errors refer to uncertainties associated with the quark-mass ratio $m_c/m_b$, CKM parameters, and higher-order perturbative corrections. This suggests that finding an asymmetry outside the range $0<{\cal A}_{X_s\gamma}<1\%$ would be a clean signal of new physics. Indeed, various extensions of the SM have been analyzed with regard to the constraints arising from the measured value of the CP asymmetry \cite{Kagan:1998bh,Wolfenstein:1994jw,Asatrian:1996as,Borzumati:1998tg,Hurth:2003vb,Blanke:2006sb,Ellis:2007kb,Altmannshofer:2009ne,Soni:2010xh,Buras:2010pi,Jung:2010ab}, and reducing the experimental error to the 1\% level is considered an important goal of future super-flavor factories.

In recent work \cite{Benzke:2010js}, we have presented a new factorization formula for inclusive radiative $B$ decays in the relevant region of large photon energy. In addition to the familiar ``direct photon contributions'', in which the photon couples to a local operator mediating the weak decay in the effective low-energy theory, novel ``resolved photon contributions'' appear. They account for the hadronic substructure of the photon, which is probed when it couples to light collinear partons. Importantly, even after integrating over the photon energy spectrum, the resolved photon contributions cannot be described using a local operator-product expansion. They give rise to first-order $\Lambda_{\rm QCD}/m_b$ corrections to the inclusive decay rate and CP asymmetry, which are proportional to $B$-meson matrix elements of non-local operators in heavy-quark effective theory (HQET). The analysis of \cite{Benzke:2010js} has shown that these non-perturbative effects lead to an irreducible uncertainty in the theoretical prediction for the CP-averaged $\bar B\to X_s\gamma$ branching ratio of about $\pm 5\%$.

An interesting feature of the resolved photon contributions is that they give rise to novel, calculable strong-interaction phases related to hard-collinear jet functions, which are convoluted with real, non-perturbative soft functions. It is interesting to explore the potential impact of these contributions on the inclusive CP asymmetries. While these effects are still of order $\Lambda_{\rm QCD}/m_b$, we show that they give the parametrically leading contribution to the $\bar B\to X_{s,d}\,\gamma$ asymmetries in the SM, and that they can strongly influence the way in which new-physics effects might show up.

Experiments measure the CP asymmetry ${\cal A}_{X_s\gamma}(E_0)$ defined with a lower cut on the photon energy, $E_\gamma\ge E_0$, with $E_0$ ranging between 1.9 and 2.2\,GeV. Detailed theoretical studies have shown that the dependence of the asymmetry on the value of $E_0$ is very mild \cite{Kagan:1998bh}. We will assume for simplicity that the cutoff can be chosen sufficiently low, such that $\Delta\equiv m_b-2E_0=\mbox{few\,$\times\Lambda_{\rm QCD}$}$ is in the perturbative domain. We will refer to the asymmetry defined with such a cut as ``partially inclusive''. In this case the direct photon contributions to the CP asymmetry can be calculated in terms of local operator matrix elements using a combined expansion in powers of $\Delta/m_b$ and $\Lambda_{\rm QCD}/\Delta$ \cite{Neubert:2004dd}. The resolved photon contributions can still not be expressed in terms of local matrix elements. However, the relevant soft functions can be simplified in this limit, as described in \cite{Benzke:2010js}. 

\begin{figure}
\begin{center}
\includegraphics[width=0.31\columnwidth]{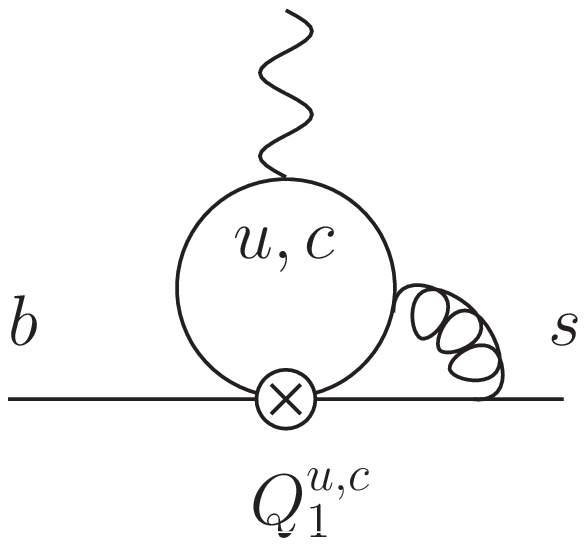} 
\includegraphics[width=0.31\columnwidth]{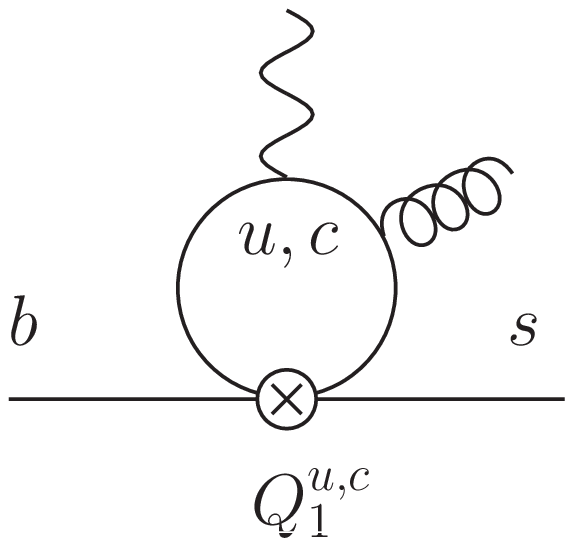} 
\includegraphics[width=0.31\columnwidth]{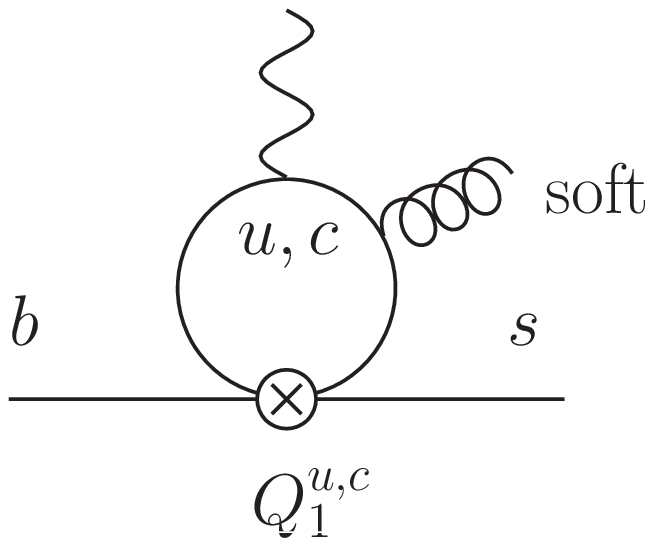} 
\caption{\label{fig:graphs} 
Penguin diagrams with virtual and real gluon emissions. The gluons in the first two graphs are highly energetic.}
\end{center}
\vspace{-6mm}
\end{figure}

The direct photon contributions to the partially inclusive CP asymmetry reduce to perturbative expressions depending on a cut parameter $\delta=\Delta/m_b$. At first non-trivial order in $\alpha_s$, one obtains \cite{Kagan:1998bh,Asatryan:2000kt}
\begin{widetext}
\begin{equation}\label{Acp_direct}
\begin{aligned}
   {\cal A}_{X_s\gamma}^{\rm dir}(E_0)
   &= \alpha_s\,\Bigg\{ \frac{40}{81}\,\mbox{Im}\,\frac{C_1}{C_{7\gamma}} 
    - \frac{8z}{9}\,\Big[ v(z) + b(z,\delta) \Big]\,
    \mbox{Im}\bigg[(1+\epsilon_s)\,\frac{C_1}{C_{7\gamma}}\bigg] 
    - \frac49\,\mbox{Im}\,\frac{C_{8g}}{C_{7\gamma}} \\
   &\hspace{1.15cm}\mbox{}+ \frac{8z}{27}\,b(z,\delta)\,
    \frac{\mbox{Im}[(1+\epsilon_s)\,C_1 C_{8g}^*]}{|C_{7\gamma}|^2} 
    + \frac{16z}{27}\,\tilde b(z,\delta)\,\bigg|\frac{C_1}{C_{7\gamma}}\bigg|^2\,
    \mbox{Im}\,\epsilon_s \Bigg\} \,,
\end{aligned}
\end{equation}
\end{widetext}
where $z=(m_c/m_b)^2$, and $\epsilon_s=(V_{ub} V_{us}^*)/(V_{tb} V_{ts}^*)=\lambda^2(i\bar\eta-\bar\rho)/[1-\lambda^2(1-\bar\rho+i\bar\eta)]+{\cal O}(\lambda^6)$ in terms of Wolfenstein parameters. The Wilson coefficients of the electro-magnetic and chromo-magnetic dipole operators in the effective weak Hamiltonian are denoted by $C_{7\gamma}$ and $C_{8g}$, while $C_1$ is the coefficient of the dominant current-current operator. Contributions from other operators are negligibly small in the SM. We suppress the scale dependence of $\alpha_s$ and $C_i$. Explicit expressions for the functions $v(z)$, $b(z,\delta)$, and $\tilde b(z,\delta)$ can be found in \cite{Kagan:1998bh,inprep}. In the SM the Wilson coefficients are real, and only terms in (\ref{Acp_direct}) proportional to the imaginary part of $\epsilon_s$ contribute. The numerically most important contributions arise from the interference of charm- and up-quark penguin graphs with virtual or real gluon emission (first two diagrams in Figure~\ref{fig:graphs}) with the leading electro-magnetic dipole amplitude. The above expression is however more general, as it holds for all new-physics models in which the dominant non-standard effects are described by additional (possibly complex) contributions to the dipole coefficients $C_{7\gamma}$ and $C_{8g}$. It would be straightforward to extend the analysis to include opposite-chirality dipole operators \cite{Kagan:1998bh}.

The above expression simplifies considerably if we adopt the power counting $m_c^2={\cal O}(m_b\Lambda_{\rm QCD})$ for the charm-quark mass and expand in powers of $z,\delta={\cal O}(\Lambda_{\rm QCD}/m_b)$. The terms proportional to $b(z,\delta)$ and $\tilde b(z,\delta)$ scale as $(\Lambda_{\rm QCD}/m_b)^2$ and can be neglected to a good approximation, while the contribution proportional to $v(z)$ contributes at first order in $\Lambda_{\rm QCD}/m_b$ and can be simplified by expanding $v(z)$ to zeroth order in $z$. This yields the approximation
\begin{eqnarray}\label{Acpdirappr}
   {\cal A}_{X_s\gamma}^{\rm dir}
   &=& \alpha_s\,\Bigg\{ \frac{40}{81}\,\mbox{Im}\frac{C_1}{C_{7\gamma}}
    - \frac49\,\mbox{Im}\frac{C_{8g}}{C_{7\gamma}} \\
   &&\qquad\mbox{}- \frac{40\Lambda_c}{9m_b}\,\mbox{Im}\bigg[(1+\epsilon_s)\,
    \frac{C_1}{C_{7\gamma}}\bigg] 
    + {\cal O}\bigg( \frac{\Lambda_{\rm QCD}^2}{m_b^2} \bigg) \Bigg\} \,, \nonumber
\end{eqnarray}
which is independent of the cutoff $E_0$. Here we have introduced the scale $\Lambda_c\sim\Lambda_{\rm QCD}$ defined as (we use $m_b=4.65$\,GeV and $m_c=1.13$\,GeV as in \cite{Benzke:2010js}) 
\begin{equation}
   \Lambda_c\equiv \frac{m_c^2}{m_b} \left( 1 - \frac25 \ln\frac{m_b}{m_c} 
    + \frac45 \ln^2\frac{m_b}{m_c} - \frac{\pi^2}{15}\right)
   \approx 0.38\,\mbox{GeV} \,.
\end{equation} 
In the SM only the last term in (\ref{Acpdirappr}) contributes, which exhibits the triple suppression by $\alpha_s$, $\mbox{Im}(\epsilon_s)\sim\lambda^2$, and $(m_c/m_b)^2\sim\Lambda_{\rm QCD}/m_b$. 

Using the factorization analysis of $\bar B\to X_s\gamma$ decay performed in \cite{Benzke:2010js}, it is possible to derive for the first time the resolved photon contributions to the partially inclusive CP asymmetry. They arise from the interference of the electro-magnetic dipole amplitude with resolved photon contributions involving up- and charm-quark penguin transitions or chromo-magnetic dipole transitions. The result can be expressed in terms of three hadronic parameters $\tilde\Lambda_{ij}$ related to convolution integrals over two soft functions denoted by $h_{17}(\omega)$ and $h_{78}^{(1)}(\omega_1,\omega_2)$. At lowest order in $\alpha_s$ and $1/m_b$ we find
\begin{eqnarray}\label{Acp_resolved}
   {\cal A}_{X_s\gamma}^{\rm res}
   &=& \frac{\pi}{m_b}\,\bigg\{
    \mbox{Im}\bigg[(1+\epsilon_s)\,\frac{C_1}{C_{7\gamma}}\bigg]\,\tilde\Lambda_{17}^c 
    - \mbox{Im}\bigg[\epsilon_s\,\frac{C_1}{C_{7\gamma}}\bigg]\,\tilde\Lambda_{17}^u \nonumber\\
   &&\hspace{0.8cm}\mbox{}+ \mbox{Im}\,\frac{C_{8g}}{C_{7\gamma}}\,
    4\pi\alpha_s\,\tilde\Lambda_{78}^{\bar B} \bigg\} \,,
\end{eqnarray}
where (omitting the scale dependence of the soft functions and $\tilde\Lambda_{ij}$ parameters)
\begin{equation}\label{Lamijdef}
\begin{aligned}
   \tilde\Lambda_{17}^u &= \frac23\,h_{17}(0) \,, \\
   \tilde\Lambda_{17}^c 
   &= \frac23 \int_{4m_c^2/m_b}^\infty\!\frac{d\omega}{\omega}\,
    f\bigg( \frac{m_c^2}{m_b\,\omega} \bigg)\,h_{17}(\omega) \,, \\
   \tilde\Lambda_{78}^{\bar B}
   &= 2\int_{-\infty}^\infty\!\frac{d\omega}{\omega}
    \left[ h_{78}^{(1)}(\omega,\omega) - h_{78}^{(1)}(\omega,0) \right] ,
\end{aligned}
\end{equation}
with 
\begin{equation}
   f(x) = 2x\ln\frac{1+\sqrt{1-4x}}{1-\sqrt{1-4x}} \,.
\end{equation}
The functions $h_{ij}$ are defined in terms of $B$-meson matrix elements of non-local operators in HQET \cite{Benzke:2010js}. In light-cone gauge $n\cdot A=0$, we have
\begin{equation}
   h_{17}(\omega) 
   = \int\frac{dt}{2\pi}\,e^{-i\omega t}\,
    \frac{\langle\bar B| \bar h(0)\,\rlap{/}{\bar n}\,i\gamma_\alpha^\perp\bar n_\beta\,
    g G^{\alpha\beta}(t\bar n)\,h(0) |\bar B\rangle}{2M_B} \,, 
\end{equation}
where $n^\mu=(1,0,0,1)$ and $\bar n^\mu=(1,0,0,-1)$ are two light-like vectors, and $h$ are effective heavy-quark fields in HQET \cite{Neubert:1993mb}. Similarly, $h_{78}^{(1)}(\omega_1,\omega_2)$ is given in terms of a matrix element of a four-quark operator \cite{inprep}. The function $h_{17}$ is an even function of $\omega$, whose normalization is equal to the HQET parameter $2\lambda_2\approx (0.5\,\mbox{GeV})^2$. No rigorous constraints are known for the function $h_{78}^{(1)}$. Note that the parameter $\tilde\Lambda_{78}^{\bar B}$ in (\ref {Acp_resolved}) depends on the flavor of the spectator quark inside the $\bar B$ meson. In the limit of SU(3) flavor symmetry, it can be shown that $\tilde\Lambda_{78}^{\bar B}=e_{\rm spec}\,\tilde\Lambda_{78}$ \cite{Benzke:2010js}, where $e_{\rm spec}$ denotes the electric charge of the spectator quark in units of $e$ ($e_{\rm spec}=2/3$ for $B^-$ and $-1/3$ for $\bar B^0$). Evaluating the hadronic matrix element of the corresponding non-local four-quark operator in the vacuum insertion approximation, we find that
\begin{equation}
   \tilde\Lambda_{78}^{\bar B} \approx e_{\rm spec}\,\tilde\Lambda_{78}
   \approx e_{\rm spec}\,\frac{2f_B^2 M_B}{9}
    \int_0^\infty\!d\omega\,\frac{\left[\phi_+^B(\omega)\right]^2}{\omega} \,,
\end{equation}
where $f_B\approx 193$\,MeV is the $B$-meson decay constant and $\phi_+^B$ its leading light-cone distribution amplitude \cite{Grozin:1996pq}.

At present, there does not exist any systematic theoretical approach to determine the hadronic parameters $\tilde\Lambda_{ij}$ from first principles. Numerical estimates must then be obtained by modeling the corresponding soft functions. Employing the models studied in \cite{Benzke:2010js}, we obtain the ranges
\begin{equation}\label{models}
\begin{aligned}
   - 330\,\mbox{MeV} < &\tilde\Lambda_{17}^u < + 525\,\mbox{MeV} \,, \\
   - 9\,\mbox{MeV} < &\tilde\Lambda_{17}^c < + 11\,\mbox{MeV} \,, \\
   17\,\mbox{MeV} < &\tilde\Lambda_{78} < 190\,\mbox{MeV} \,.
\end{aligned}
\end{equation}
All three estimates are very uncertain, but we observe that $\tilde\Lambda_{17}^u$ and $\tilde\Lambda_{78}$ are expected to be of order $\Lambda_{\rm QCD}$. The slight preference for positive values of $\tilde\Lambda_{17}^q$ is due to the normalization constraint on the function $h_{17}$ mentioned above. Note that in the formal limit $m_c\to m_u=0$ the values of $\tilde\Lambda_{17}^u$ and $\tilde\Lambda_{17}^c$ coincide. However, we predict a strong GIM violation owing to the fact that the integral in the second relation in (\ref{Lamijdef}) starts at $4m_c^2/m_b\approx 1.1$\,GeV, at which the soft function is expected to take already rather small values, since it is governed by non-perturbative dynamics and must vanish for $\omega\to\infty$.

The complete theoretical result for the partially inclusive CP asymmetry in $\bar B\to X_s\gamma$ decay is obtained by adding the direct and resolved contributions (\ref{Acp_direct}) and (\ref{Acp_resolved}). In order to understand better the structure of the result, we now present some approximate formulae obtained by using expression (\ref{Acpdirappr}) for the direct contribution. Our numerical results will always be derived using the exact expression. For the CP asymmetry in the SM, we obtain
\begin{equation}\label{AcpSMtot}
\begin{aligned}
   {\cal A}_{X_s\gamma}^{\rm SM}
   &\approx \pi\,\bigg|\frac{C_1}{C_{7\gamma}}\bigg|\,\mbox{Im}\,\epsilon_s\,
    \bigg( \frac{\tilde\Lambda_{17}^u-\tilde\Lambda_{17}^c}{m_b}
    + \frac{40\alpha_s}{9\pi}\,\frac{\Lambda_c}{m_b} \bigg) \\
   &= \left( 1.15\times\frac{\tilde\Lambda_{17}^u-\tilde\Lambda_{17}^c}{300\,\mbox{MeV}}
    + 0.71 \right) \% \,.
\end{aligned}
\end{equation}
We fix the photon cut at $E_0=1.9$\,GeV and use $\lambda=0.2254$, $\bar\rho=0.144$, $\bar\eta=0.342$ for the Wolfenstein parameters. Our choice $\mu=2$\,GeV for the factorization scale (for which $\alpha_s(\mu)=0.307$, and $C_1(\mu)=1.204$, $C_{7\gamma}(\mu)=-0.381$, $C_{8g}(\mu)=-0.175$ at leading order) is motivated by the fact that the strong phases required for a non-zero CP asymmetry arise either from GIM violations related to charm-quark loops (for which $\mu\sim 2m_c$), or from cut hard-collinear propagators (for which $\mu\sim\sqrt{m_b\Lambda_{\rm QCD}}$). In the resolved photon term we keep the contribution of $\tilde\Lambda_{17}^c$ in order to make explicit that the CP asymmetry vanishes in the formal limit $m_c=m_u$ due to the GIM mechanism. In practice, however, this contribution can be safely neglected. The dominant contribution arises from the up-quark penguin graph with emission of a soft gluon, in which the quark loop is probed at a light-like distance away from the heavy quarks (last diagram in Figure~\ref{fig:graphs}). We do not show the dependences of our results on variations of the input parameters, which are much smaller than the uncertainties associated with the resolved photon contributions. Our central value 0.71\% for the direct photon term is larger than that obtained in \cite{Hurth:2003dk} since we use smaller values for $\mu$ and $m_c$.

The resolved photon term proportional to $\tilde\Lambda_{17}^u$ in (\ref{AcpSMtot}) is parametrically larger than the direct photon term, which contains an additional $\alpha_s$ suppression. Numerically, this term dominates as long as $|\tilde\Lambda_{17}^u|$ is larger than about 200\,MeV. Using the model estimates shown in (\ref{models}) we find the range $-0.6\%<{\cal A}_{X_s\gamma}^{\rm SM}<2.8\%$, which covers most of the experimentally allowed range (\ref{data}). Only a value of the asymmetry below $-2\%$ could be interpreted as a sign of new physics, as in this case $\tilde\Lambda_{17}^u<-700$\,MeV would be much larger in magnitude than our model expectations. 

In extensions of the SM, in which the dipole coefficients $C_{7\gamma}$ and $C_{8g}$ receive new CP-violating contributions, additional terms arise. Using the approximation (\ref{Acpdirappr}), we find
\begin{eqnarray}
   \frac{{\cal A}_{X_s\gamma}}{\pi}
   &\approx& \left[ \left( \frac{40}{81} - \frac{40}{9}\,\frac{\Lambda_c}{m_b} \right) 
    \frac{\alpha_s}{\pi} 
    + \frac{\tilde\Lambda_{17}^c}{m_b} \right] \mbox{Im}\,\frac{C_1}{C_{7\gamma}} \nonumber\\
   &&\mbox{}- \left( \frac{4\alpha_s}{9\pi} - 4\pi\alpha_s\,e_{\rm spec}\,
    \frac{\tilde\Lambda_{78}}{m_b} \right) \mbox{Im}\,\frac{C_{8g}}{C_{7\gamma}} \\
   &&\mbox{}- \left( \frac{\tilde\Lambda_{17}^u - \tilde\Lambda_{17}^c}{m_b}
    + \frac{40}{9}\,\frac{\Lambda_c}{m_b}\,\frac{\alpha_s}{\pi} \right)
    \mbox{Im}\bigg(\epsilon_s\,\frac{C_1}{C_{7\gamma}}\bigg) \,. \nonumber
\end{eqnarray}
For the first two terms the resolved photon contributions give rise to power corrections to the direct photon terms, which are numerically significant since $\alpha_s/\pi\sim\Lambda_{\rm QCD}/m_b$. For the third term, which is the only one present in the SM, the resolved photon contribution is likely to be more important than the direct photon term.

To illustrate the impact of the resolved photon terms, we consider the class of new-physics models in which the dominant non-standard effects are encoded in the values of the dipole operators, which we parameterize in the form $C_{7\gamma}/C_1=(C_{7\gamma}/C_1)^{\rm SM}\,r_7\,e^{i\theta_7}$ and $C_{8g}/C_1=(C_{8g}/C_1)^{\rm SM}\,r_8\,e^{i\theta_8}$. Using that $\mbox{arg}(-\epsilon_s)\approx -\gamma$ in the SM, we then obtain for our default parameters
\begin{equation}
\begin{aligned}
   {\cal A}_{X_s\gamma}~[\%]
   &= \left( 10.12 + 2.14\times\frac{\tilde\Lambda_{17}^c}{10\,\mbox{MeV}} \right) 
    \frac{1}{r_7}\,\sin\theta_7 \\
   &\hspace{-1cm}\mbox{}
    + \left( 1.26\times\frac{\tilde\Lambda_{17}^u-\tilde\Lambda_{17}^c}{300\,\mbox{MeV}}
    + 0.74 \right) \frac{1}{r_7}\,\sin(\gamma+\theta_7) \\
   &\hspace{-1cm}\mbox{}+ \left( 6.27 - 11.98\,e_{\rm spec}\times
    \frac{\tilde\Lambda_{78}}{100\,\mbox{MeV}} \right) \frac{r_8}{r_7}\,\sin(\theta_7-\theta_8) \\
   &\hspace{-1cm}\mbox{}+ 0.18\,\frac{r_8}{r_7^2}\,\sin\theta_8 
    + \frac{0.037}{r_7^2}\,\sin\gamma - 0.004\,\frac{r_8}{r_7^2}\,\sin(\gamma+\theta_8) \,. 
\end{aligned}
\end{equation}
For the flavor-averaged CP asymmetry, $\frac12\big({\cal A}_{X_s^-\gamma}\!+{\cal A}_{X_s^0\gamma}\big)$, we must replace $e_{\rm spec}\to\frac16$. Then the resolved photon contributions are subdominant except for the second term, which is already present in the SM. In principle very large asymmetries are possible from the first and third terms \cite{Kagan:1998bh}, which however are already ruled out by the data. Once we assume that the effects of new physics are at most a few percent in magnitude, it will be difficult to disentangle them from the hadronic uncertainty due to the $\tilde\Lambda_{17}^u$ parameter.

A non-trivial feature of our analysis is that the resolved photon contributions induce a flavor-dependent term in the CP asymmetry already at order $\Lambda_{\rm QCD}/m_b$. In the SM such effects are suppressed, compared with (\ref{AcpSMtot}), by at least one additional factor of $\Lambda_{\rm QCD}/m_b$ and are thus bound to be negligible. We thus propose a future precision measurement of the CP asymmetry difference
\begin{equation}
\begin{aligned}
   \Delta{\cal A}_{X_s\gamma} 
   &\equiv {\cal A}_{X_s^-\gamma} - {\cal A}_{X_s^0\gamma}
    \approx 4\pi^2\alpha_s\,\frac{\tilde\Lambda_{78}}{m_b}\,
    \mbox{Im}\,\frac{C_{8g}}{C_{7\gamma}} \\
   &\approx 12\% \times \frac{\tilde\Lambda_{78}}{100\,\mbox{MeV}}\,
    \frac{r_8}{r_7}\,\sin(\theta_8-\theta_7)
\end{aligned}
\end{equation}
as a sensitive probe for flavor physics beyond the SM. Even though it will be difficult to determine the value of the hadronic parameter $\tilde\Lambda_{78}$ with any reasonable accuracy, we observe that if either the electro-magnetic or the chromo-magnetic dipole coefficients (or both) receive a sizable CP-violating new-physics phase, the difference $\Delta{\cal A}_{X_s\gamma}$ can easily reach the level of 10\% in magnitude. It is important in this context that the Wilson coefficient of the chromo-magnetic operator can be much enhanced with regard to its SM value, so that $r_8/r_7\sim\mbox{a few}$ is possible \cite{Kagan:1998bh}.

All of the above expressions also apply to the CP asymmetry in $\bar B\to X_d\gamma$ decay, once we replace $\epsilon_s$ by $\epsilon_d=(V_{ub} V_{ud}^*)/(V_{tb} V_{td}^*)=(\bar\rho-i\bar\eta)/(1-\bar\rho+i\bar\eta)$. As a result, the CP asymmetry for $\bar B\to X_d\gamma$ decay in the SM differs by that in $\bar B\to X_s\gamma$ decay by a factor $\mbox{Im}(\epsilon_d)/\mbox{Im}(\epsilon_s)\approx -22$. With the parameter values shown in (\ref{models}), we obtain the range $-62\%<{\cal A}_{X_d\gamma}^{\rm SM}<14\%$. It is worth emphasizing that in the endpoint region of large photon energy there do exist contributions to the photon energy spectrum which could affect the CP asymmetries in the decays $\bar B\to X_s\gamma$ and $\bar B\to X_d\gamma$ in a way different from the above rescaling. They are sensitive to four-quark soft functions with flavor content $\bar b s\bar s b$ and $\bar b d\bar d b$, respectively, whose matrix elements between $B_d$ meson states can be different even in the flavor SU(3) limit. However, as shown in \cite{Benzke:2010js} these effects integrate to zero in the partially inclusive CP asymmetries at order $\Lambda_{\rm QCD}/m_b$. As a result, the untagged CP asymmetry for $\bar B\to X_{s+d}\,\gamma$ decays vanishes in the SM (up to tiny U-spin breaking corrections \cite{Soares:1991te,Kagan:1998bh,Hurth:2001yb}) even after the resolved photon terms are taken into account. 

{\em Acknowledgements:\/}
We are grateful to Tobias Hurth for useful discussions. The research of M.B.\ and M.N.\ is supported by BMBF grant 05H09UME and by the Research Centre of Excellence {\em Elementary Forces and Mathematical Foundations}. The work of G.P.\ is supported by the DOE grant DE-FG02-90ER40560.

\end{document}